\documentclass[numreferences]{crckbked}
\usepackage{graphicx}
\usepackage{klucite}
\begin{document}


\def\med{{1\ov 2}}
\def\hepth#1{ {\tt hep-th/#1}}
\def\mathph#1{ {\tt math-ph/#1}}
\def\alggeom#1{ {\tt alg-geom/#1}}
\def\res{{\rm res}}
\def\min{{\rm min}}
\def\max{{\rm max}}
\def\inv{^{-1}}
\def\bd{\begin{description}}
\def\ed{\end{description}}
\def\etal{{\it et al. }}
\def\ie{{\it i.e. }}
\def\eg{{\it e.g. }}
\def\Cf{{\it Cf.\ }}
\def\cf{{\it cf.\ }}
\def\be{\begin{equation}}
\def\ee{\end{equation}}
\def\bes{\begin{equation*}}
\def\ees{\end{equation*}}
\def\beqa{\begin{eqnarray}}
\def\beqas{\begin{eqnarray*}}
\def\eeqa{\end{eqnarray}}
\def\eeqas{\end{eqnarray*}}
\def\bea{\begin{eqnarray}}
\def\eea{\end{eqnarray}}
\def\tr{{\rm tr}}
\def\Tr{{\rm Tr}}
\def\d{\partial}
\def\ov{\over}
\def\pder#1#2{\frac{\partial #1}{\partial #2}}
\def\der#1#2{\frac{d #1}{d #2}}
\def\ppder#1#2#3{\frac{\partial^2 #1}{\partial #2\partial #3}}
\def\dpder#1#2{\frac{\partial^2 #1}{\partial #2 ^2 }}
\def\theequation{\thesection.\arabic{equation}}
\def\cdotsk{\!\cdot\!}
\def\cdotsh{\!\cdot}
\newcommand{\comps}{\mathbb{C}}
\newcommand{\reals}{\mathbb{R}}
\newcommand{\integs}{\mathbb{Z}}


\def\a{\alpha}
\def\b{\beta}
\def\e{\epsilon}
\def\g{\gamma}
\def\G{\Gamma}
\def\p{\phi}
\def\r{\rho}
\def\d{\delta}
\def\l{\lambda}
\def\m{\mu}
\def\om{\omega}
\def\s{\sigma}
\def\L{\Lambda}
\def\P{\Phi}


\def\cF{\cal F}
\def\cH{\cal H}
\def\cL{\cal L}
\def\cM{\cal M}
\def\cN{\cal N}
\def\cO{\cal O}
\def\cT{\cal T}
\def\cW{\cal W}


\begin{opening}
\title{Integrable hierarchies in Donaldson-Witten \\ and Seiberg-Witten 
theories \thanks{Invited Talk at the NATO Advanced Research Workshop {\em
Integrable Hierarchies and Modern Physical Theories}, University of Illinois,
Chicago, 22th-26th July 2000. Preprint numbers: HUTP-00/A040, US-FT/15-00}}

\author{Jos\'e D. \surname{Edelstein} \thanks{\tt edels@lorentz.harvard.edu}}

\institute{Lyman Laboratory of Physics \\ Harvard University \\ Cambridge, MA 
02138, USA}

\author{Marta \surname{G\'omez--Reino} \thanks{\tt marta@fpaxp1.usc.es}}

\institute{Departamento de F\'{\i}sica de Part\'{\i}culas \\ Universidade de 
San\-tia\-go de Compostela \\ E-15706 Santiago de Compostela, Spain}

\begin{abstract}
We review various aspects of integrable hierarchies appearing in ${\cal N}=2$
supersymmetric gauge theories. In particular, we show that the blowup function
in Donaldson--Witten theory, up to a redefinition of the fast times, is a 
$\tau$--function for a $g$-gap solution of the KdV hierarchy. In the case
of four-manifolds of simple type, instead, the blowup function becomes a 
$\tau$--function corresponding to a multisoliton solution. We obtain a new 
expression for the contact terms that links these results to the Whitham 
hierarchy formulation of Seiberg--Witten theories.
\end{abstract}
\end{opening}

\section{Introduction}

The Seiberg--Witten ansatz for the low-energy effective action of ${\cal N}
= 2$ super Yang--Mills theories \cite{SeWi,SeWi2} stands out as the only exact
solution that is known at present in four-dimensional Quantum Field Theory. 
The quantum moduli space of vacua of the theory ${\cal M}$ is identified 
with the moduli space of an auxiliary hyperelliptic complex curve $\Sigma$,
in such a way that there is a selected meromorphic differential $dS$ --that
induces a special geometry on $\Sigma$--, whose periods give the 
spectrum of BPS states. Interestingly enough, this solution has been shown to 
display remarkable nonperturbative phenomena such as quark confinement by
monopole condensation, when a mass term that breaks supersymmetry
down to ${\cal N} = 1$ is included. In the sake of definiteness and clarity,
we will restrict our discussion to the case of ${\cal N} = 2$ supersymmetric
theories with gauge group $SU(N)$ and without matter content 
\cite{KlLeThYa,ArFa}, though many of the results reviewed here can be 
appropriately generalized in different directions.

It was presently realized that the Seiberg--Witten solution could be 
reformulated, in an elegant and useful way, in terms of classical finite-gap 
integrable systems, $dS$ being a solution of their averaged (Whitham) 
dynamics \cite{GKMMM}. The spectral curve of the integrable system $\G$ is 
identified with the auxiliary hyperelliptic curve $\Sigma$, and the effective
prepotential of the supersymmetric gauge theory ${\cal F}(a^i,\Lambda)$ 
turns out to be given by the logarithm of the quasiclassical $\tau$--function.
In the case of pure gauge theory, for example, the corresponding integrable
system is the periodic Toda chain \cite{MaWa,NaTa}. The 
quassiclasical Whitham hierarchy associated to adiabatic deformations of 
the integrable system naturally endowes moduli which, instead of being 
local invariants, evolve with respect to the {\em slow times} $T_n$ 
\cite{Kr}. The upshot of this formalism is a prepotential also depending on 
these new variables, ${\cal F}(a^i,T_n)$ \cite{ItMo,ItMo2}. This prepotential
is a deformation of the former in the sense that, roughly speaking, if we 
put $T_1 = \Lambda$ and $T_{n>1} = 0$, the Seiberg--Witten prepotential 
is recovered.

The Whitham dynamics can be thought of as a sort of generalization of the
Renormalization Group flow \cite{GoMaMiMo}. Aside from its intrinsic
formal interest, it governs a relevant family of deformations 
of the Seiberg--Witten solution. In fact, it turns out that the slow times are 
{\em dual} to homogeneous combinations of higher Casimir operators, this 
revealing that they are the appropriate variables to be promoted to spurion 
superfields if one is interested in softly broken ${\cal N} = 2$ supersymmetry
by means of higher than quadratic ${\cal N} = 0$ perturbations \cite{EdMaMa}.
The lowest slow time can be identified with the quantum dynamical scale
$\Lambda$ of the supersymmetric gauge theory, and its uses as a spurion has
been extensively studied in \cite{AGDiKoMa,AGMa}. Moreover, this formalism 
is also very fruitful when restricted to the original Seiberg--Witten 
variables; new equations arise that provide a powerful technique allowing to
compute interesting quantities in the infrared such as instanton corrections
up to arbitrary order \cite{EdMaMa,EdGRMa}, and the strength of the coupling
among different {\em magnetic photons} at the monopole singularities 
\cite{EdMa}. The Whitham hierarchy, and the details of its connection to 
four-dimensional ${\cal N}=2$ supersymmetric gauge theories, have been already
the focus of many comprehensive reviews \cite{rev1,rev2,rev3,rev4} and books
\cite{book1,book2}.

There is another context, definitely less explored, where four-dimensio\-nal
${\cal N} = 2$ supersymmetric theories get involved with integrability. 
It is the case of topological field theories built out
of twisted versions of ${\cal N} = 2$ supersymmetric gauge theories 
\cite{Wi88}. Soon after the appearance of Ref.\cite{SeWi,SeWi2}, it was 
realized that the Seiberg--Witten solution may be used to compute Donaldson
invariants of a four-manifold $X$ by counting solutions of Abelian monopole
equations \cite{Wi94}. Furthermore, for manifolds with $b_2^+(X) = 1$, a 
nontrivial contribution comes from each point of the Coulomb branch so that 
the path integral used as a generating function for Donaldson invariants 
acquires the form of a sort of integral over ${\cal M}$ (the so-called 
$u$-plane integral \cite{MoWi,MaMo}). The Whitham hierarchy formalism has been 
shown to provide an adequate conceptual framework to study several aspects 
of the behavior of the $u$-plane integral under blowup of a point $p \in X$.
The contact terms corresponding to pairs of observables, for example, can 
be written as derivatives of the prepotential with respect to $T_n$ variables
\cite{MaMo2,Ta,Ma}. On the other hand, these terms can be derived from
the blowup function \cite{LoNeSh,LoNeSh2}, which is nothing but a factor 
appearing in the so-called {\em blowup formula} \cite{FiSt,GoZa} that 
relates the path integral in $X$ with the one performed in the blownup 
manifold $\hat X$. Furthermore, the blowup formula itself involves in a 
rather direct way the underlying integrable structure of the low-energy 
effective theory.

\begin{figure}
\centerline{\includegraphics[height=7.6cm]{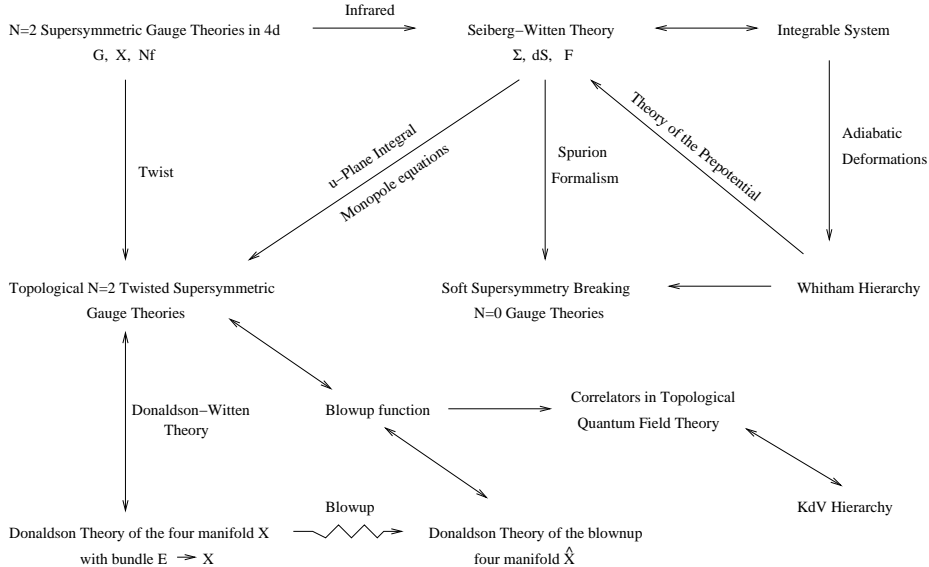}}
\caption{Schematic picture of the different connections between
four-dimensional ${\cal N}=2$ supersymmetric gauge theories and integrable 
hierarchies.}
\label{funo}
\end{figure}

In previous remarks we have discussed a connection between integrability 
and four-dimensional supersymmetric gauge theories that heavily relies in 
the identification of the relevant geometrical data, and the uses of 
Whitham theory to study adiabatic deformations thereof. A rough pattern of 
these interrelations is given in Fig.\ref{funo}. In all known cases --that 
is, for any gauge group and matter content--, the relevant integrable system
has a finite number of degrees of freedom. There is, however, a different 
way in which integrability enters into the game of twisted ${\cal N}=2$ 
supersymmetric gauge theories. It emerges from a detailed analysis of the 
blowup formula under the light of the theory of hyperelliptic Kleinian 
functions. The main consequence of such investigation is that the blowup 
function in Donaldson--Witten theory, up to a redefinition of the {\em
fast} times, is a $\tau$--function for a $g$-gap solution of the KdV 
hierarchy \cite{EdGRMar}. For manifolds of the so-called {\em simple type}, 
the blowup function becomes the $\tau$--function of a
{\em multisoliton} solution of the hierarchy. As a corollary, the correlation
functions involving the exceptional divisor on the blownup manifold are
governed by the KdV hierarchy, this giving an intriguing connection with
two-dimensional topological gravity \cite{Wi91}. The uses of the theory of 
hyperelliptic Kleinian functions also provides powerful techniques that 
further enhance our knowledge of contact terms and blowup functions. It is 
our purpose in this talk to present a short survey comprising these
latest results.

\section{Brief overview of Donaldson-Witten theory}

Let $X$ be a smooth, compact, oriented (for simplicity, we also assume 
that it is simply connected) four-manifold with Riemannian metric $g$, and
let $E \to X$ be an $SU(2)$ bundle. The degree $s$ Donaldson polynomial,
defined on the homology of $X$ with rational coefficients, is given by 
\cite{DoKr}
\be
{\cal D}_E(p,S) = \sum_{2n+4t=s} d_{n,t}~S^n~p^t ~,
\label{donpol}
\ee
where $p\in H_0(X,\dQ)$, $S \in H_2(X,\dQ)$, and $s$ is the dimension of 
the moduli space of instantons on $E$. The numbers $d_{n,t}$ are precisely 
given in terms of intersection theory on this moduli space \cite{Do}.

It is useful to organize these polynomials within a generating function, by
summing over all topological types of the bundle $E$ with fixed second
Stiefel--Whitney class, $w_2(E)$,
\be
\Phi_X(p,S) = \sum_{n,t\geq 0} d_{n,t}~\frac{S^n}{n!}~\frac{p^t}{t!} ~.
\label{genfun}
\ee
The remarkable result obtained a decade ago by Witten \cite{Wi88} is that 
$\Phi_X(p,S)$ is the generating function for the correlators of observables 
in a twisted version of ${\cal N} = 2$ super Yang--Mills theory,
\be
\Phi_X(p,S) = Z_X(p,S) = \Big\langle \exp\bigl[{p\over 2}{\rm Tr}\phi^2 + 
{1\over 2}\int_{S}{\rm Tr}(\phi F) + \cdots\bigr] \Big\rangle_X ~.
\label{witten88}
\ee
Here, $\phi$ is the scalar field that belongs to the ${\cal N} = 2$ vector 
multiplet and $F$ is the Yang--Mills field strength. If $b_2^+(X)>1$, $\Phi_X$ 
is independent of $g$, thus defining topological invariants of the 
four-manifold $X$.

For gauge group $SU(N)$, there is a family of $N-1$ fundamental observables
(that is, local BRST-invariant operators), ${\cal O}_k$,
\be
{\cal O}_k = {1\over k}{\rm Tr \phi}^k + \cdots ~,
\label{fundobs}
\ee
--whose vacuum expectation values, $u_k = \langle{\cal O}_k\rangle$, are
gauge invariant coordinates in $\cM$ classifying the various vacua of the 
theory--, as well as $b_2(X)(N-1)$ topological descendants thereof, 
$I_k(S_i)$, 
\be
I_k(S_i) = {1 \over k} \int_{S_i} {\rm Tr}(\phi^{k-1}F) + \cdots ~,
\label{descobs}
\ee
where the dots stand for lower powers of $\phi$ in (\ref{fundobs}), and 
superpartner contributions in (\ref{descobs}). So, the basic problem in 
Donaldson--Witten theory is to compute the generating function
\be
Z_X(p_k,S_i)=\Big\langle \exp\bigl[ \sum_{k=2}^{N} (p_k {\cal O}_k + 
\sum_{i=1}^{b_2(X)} f_{k,i} I_k(S_i)) \bigr]\Big\rangle_X ~.
\label{DonWit}
\ee
Being metric independent, one can consider a uniparametric family
of metrics $g_t = t^2 g_0$, with fixed $g_0$, and focus on the limiting
cases $t \to 0$ and $t \to \infty$. In the former case, the topological
quantum field theory is in the ultraviolet, and thus weakly coupled, so
perturbation theory is reliable. Conversely, when $t\to\infty$ the correlation
functions result from the infrared behavior of the theory, picking up
contributions from every point in the moduli space of vacua. This amounts to
an integral over the $u$-plane \cite{MoWi,MaMo,MaMo2}.

For {\em simple type} four-manifolds, the only non-vanishing contributions
to this integral come from the maximal singularities of $\cM$. The 
infrared dynamics at those points is that of a weakly coupled theory of 
Abelian gauge fields and monopoles \cite{SeWi,SeWi2,DoSh}. Hence, roughly 
speaking, Donaldson invariants can be computed in such cases just by 
counting monopole solutions \cite{Wi94}.

The $u$-plane integral is performed by means of the Seiberg-Witten solution
\cite{MoWi,MaMo,MaMo2}. For each descent observable of the microscopic theory, 
$I_k(S_i)$, there is a corresponding low energy operator $\tilde I_k(S_i)$.
However, it is not generically true that the product of two or more of 
these operators map to the analog product in the infrared description. 
{\em Contact terms} appear whenever any pair of the supporting two-cycles 
intersect \cite{Wi95,MoWi},
\be
I_k(S_i) I_l(S_j) \longrightarrow \tilde I_k(S_i) \tilde I_l(S_j) + 
{\cT}_{k,l}(S_i \cap S_j) ~.
\label{conterm}
\ee
These terms are not deducible from the Seiberg-Witten solution. Their explicit
form can be derived from the so-called blowup function \cite{LoNeSh,LoNeSh2}
by requiring both their duality invariance and semiclassical vanishment
\cite{MoWi},
\be
{\cT}_{k,l} = - {1 \over 2 \pi i }\partial_{\tau_{ij}} \log 
\Theta[\vec\Delta,\vec 0] (0|\tau){\partial u_k \over \partial a^i}{\partial 
u_l \over \partial a^j} ~,
\label{confin}
\ee
where $\vec\Delta = (1/2,\dots,1/2)$. We will analyze in what follows the
case of four-manifolds with $b_2^+(X)=1$.

Consider now the four-manifold $\hat X$, obtained from $X$ by blowing up 
a point $p$, $\hat X = {\rm Bl}_{p}(X)$. This means that there is a map 
$\pi:\hat X \to X$ that is the identity everywhere except at $B = 
\pi^{-1}(p)$, where $B \in H_2(\hat X)$ such that $B^2 = -1$. $B$ is called 
the class of the exceptional divisor. The homology of the blownup manifold 
is the direct sum $H_2(\hat X) = H_2(X) \oplus \dZ \cdot B$. Thus, the 
twisted theory in $\hat X$ has additional descent observables $I_k(B)$ that 
must be included in the generating function. It is also possible to have a 
non-Abelian magnetic flux $\vec\beta$ through $B$ of the form $\beta^i = 
(C^{-1})^i_{\,j}~n^j$, where the $n^j$ are arbitrary integers, and 
$(C^{-1})^i_{\, j}$ is the inverse of the Cartan matrix \cite{MaMo2}. The 
generating function of the twisted theory in the blownup manifold then reads
\be
Z_{\hat X,\vec\beta}(p_k,S_i,B) = \biggl\langle \exp\bigl[ \sum_{k=2}^{N} 
(p_k {\cal O}_k + t_k I_k(B) + \sum_{i=1}^{b_2(X)} f_{k,i} I_k(S_i)) \bigr] 
\biggr\rangle_{\widehat{X},\vec\beta} ~.
\label{genfun2}
\ee
The Donaldson invariants of $\hat X$ are related to those of $X$, at least
in the case in which the exceptional divisor has a small area. This is
described by the so-called blowup formula \cite{FiSt,GoZa}. It is reflected
in the $u$-plane integral, through the following relation between both
generating functions:
\be
Z_{\hat X,\vec\beta}(p_k,S_i) = \biggl\langle \exp\bigl[ \sum_{k=2}^{N} 
(p_k {\cal O}_k + \sum_{i=1}^{b_2(X)} f_{k,i} I_k(S_i))\bigr]
~\tau_{\vec\beta}(t_k|{\cal O}_k) \biggr\rangle_X ~,
\label{blowupform}
\ee
where $\tau_{\vec\beta}(t_k|{\cal O}_k)$ is the above mentioned {\em blowup
function}. From the point of view of the original manifold, it should be a
punctual defect. Thus, it is natural to expect it to be an infinite series
of local operators \cite{MaMo2},
\be
\tau_{\vec\beta}(t_k|{\cal O}_k) = \sum_{\vec n \in Z_+^{N-1}} t^{\vec n}
{\cal B}_{\vec n,\vec\beta} ({\cal O}_2, \dots, {\cal O}_N) ~,
\label{infser}
\ee
where $\vec n=(n_2, \cdots, n_N)$, ~$t^{\vec n} \equiv t_2^{n_2} \cdots
t_N^{n_N}$, and the $n$-th order term comes from those $\vec n$ with 
$|\vec n| \equiv \sum_i n_i = n$. By means of the $u$-plane integral, the
blowup function can be written as \cite{MoWi,MaMo,MaMo2}
\be
\tau_{\vec\beta}(t_i|u_k) = {\rm e}^{-\sum_{k,l}t_k t_l{\cal T}_{k,l}}
{\Theta[\vec\Delta,\vec\beta](\vec\xi|\tau) \over \Theta[\vec\Delta,\vec 0] 
(0|\tau)} ~,
\label{blowupfunction}
\ee
where 
\be
\xi_i= \sum_{k=2}^N  {t_k \over 2\pi}{\partial u_k \over \partial a^i} ~.
\label{vector}
\ee
We shall consider in this talk the case in which there is no non-Abelian
magnetic flux through the exceptional divisor, $\vec\beta = \vec 0$. In such
a case, the quadratic contribution to the blowup function vanishes \cite{Ma},
\be
{\cal B}_{\vec n,\vec 0} ({\cal O}_2, \dots, {\cal O}_N) = 0 ~, ~~~~~
\mbox{for} ~|\vec n| = 2 ~.
\label{cuadnull}
\ee
In $SU(2)$, for example, the blowup function is nothing but an elliptic 
$\sigma$--function \cite{FiSt}. For higher rank theories, the answer should 
be some hyperelliptic generalization of the $\sigma$--function, as far 
as the blowup function is defined as a quotient of $\Theta$--functions with
a prefactor that renders it to be duality invariant; precisely the same
features that characterize a $\sigma$--function.

\section{Geometrical detour: Kleinian functions}

The theory of hyperelliptic Kleinian functions was developed one century 
ago by Baker \cite{Ba1,Ba2,Ba3} and Bolza \cite{Bo1,Bo2,Bo3,Bo4}, among 
others. A comprehensive modern survey is presented in Ref.\cite{BuEnLe}.
We shall introduce in this section some of the algebro--geometric
ingredients that will be relevant in the remainder of the talk. Let 
$\Sigma$ be a hyperelliptic curve of genus $g$, which we write in the 
{\em even} form,
\be
y^2 = f(x) = \sum_{i=0}^{2g+2}\lambda_i x^i = Q(x) R(x) ~.
\label{thecurve}
\ee
The last expression above provides a factorization of $f(x)$ in two
polynomials of degree $g+1$. We will eventually be interested in the
Seiberg--Witten setup, where
\begin{subequation}[alph]
\begin{eqnarray}
Q_{SW}(x) = P_{N}(x) - 2\Lambda^N ~, ~ & ~ & ~ R_{SW}(x) = P_{N}(x) +
2\Lambda^N ~, \\
P_N(x) = x^{N} & - & \sum_{k=2}^N u_kx^{N-k} ~.
\end{eqnarray}
\end{subequation}
Consider a symplectic basis of homology cycles, $A^i,B_i\in H_1(\Sigma,\dZ)$,
and a canonical basis of Abelian differentials of the first kind, $dv_k = 
x^{g-k} dx/y$, whose periods are
\be
A^i{_k} = {1\over 2\pi i} \oint_{A^i} dv_k ~, ~~~~~~ B_{ik} ={1\over 2\pi i}
\oint_{B_i} dv_k ~.
\label{pers}
\ee
It is well known that even and non-singular half-integer 
characteristics, $[\vec\alpha,\vec\beta]$, are in one to one correspondence 
with the different factorizations of $f(x)$ as a product of $Q(x)$ and $R(x)$.
In particular, for the Seiberg--Witten setup, it is precisely
$[\vec\Delta,\vec 0]$, the characteristic appearing in the blowup function
when there is no non-Abelian magnetic flux through $B$.
Now, in order to construct hyperelliptic $\sigma$--functions, it is also 
necessary to introduce a canonical basis of Abelian differentials of the 
second kind. This can be done by means of a Weierstrass polynomial, a 
function of two variables $F(x_1,x_2)$ which is at most of degree $g+1$
both in $x_1$ and $x_2$, and satisfies the following conditions:
\begin{subequation}[alph]
\begin{eqnarray}
F(x_1,x_2) = F(x_2,x_1) ~, ~~~ & ~ & ~~~ F(x_1,x_1) = 2 f(x_1) ~, \\
\Bigl({\partial F(x_1,x_2) \over \partial x_1} \Bigr)_{x_1=x_2} & = &
f'(x_2) ~.
\end{eqnarray}
\end{subequation}
Indeed, the following identity \cite{BuEnLe}
\be
d\vec v(x_1) \cdot d\vec r(x_2) = - {\partial \over \partial x_2} \Bigl( 
{y_2 \over x_1-x_2}\Bigr) {dx_1 \over 2 y_1} \, dx_2 + {F(x_1,x_2) \over 
4(x_1-x_2)^2} {dx_1 \over y_1} {dx_2 \over y_2} ~,
\label{identity}
\ee
implicitely defines such a basis, $dr^k(x_2)$. An example of a Weierstrass
polynomial that uses the factorization of the curve given above is
\be
F(x_1,x_2) = Q(x_1) R(x_2) + Q(x_2) R(x_1) ~.
\label{bolzapol}
\ee
The corresponding basis of Abelian differentials of the second kind, in the
Seiberg--Witten setup, reads
\be
dr^j = {1\over 2} P'_{j}(x) P_{N}(x) {dx \over y} ~, ~~~ j=1, \cdots, N-1 ~.
\label{drzero}
\ee
Two different Weierstrass polynomials are always seen to be related by
\be
F(x_1,x_2) - {\hat F}(x_1,x_2) = 4 (x_1-x_2)^2 \sum_{i,j=1}^g d_{ij} 
~x_1^{g-i} ~x_2^{g-j} ~,
\label{twodiff}
\ee
where $d_{ij}$ is symmetric in $i, j$. Correspondingly, the relation
between the two basis is
\be
dr^j = \widehat dr^j + \sum_{k=1}^g d_{jk} dv_{k} ~.
\label{reldr}
\ee
Given a basis of Abelian differentials of the second kind, one can define
the analog of elliptic $\eta$--periods through
\be
\eta^{ki} = - {1\over 2\pi i} \oint_{A^i} dr^k ~,~~~~~~~~~~~~~ 
{\eta'^k_{\,\,\ i}} = -{1\over 2\pi i}\oint_{B_i}dr^k ~.
\label{etas}
\ee
Thus, the hyperelliptic $\sigma$-function with even and non-singular 
characteristic can be defined as \cite{Bo1,Bo2,Bo3,Bo4}
\be
\sigma^{F}[\vec\alpha,\vec\beta] (\vec v) = \exp\{v_i \kappa^{il} v_l\}
{\Theta[\vec\alpha,\vec\beta] ((2\pi i)^{-1}v_l {(A^{-1})^l_{\,\ i}}|\tau) 
\over \Theta[\vec\alpha,\vec\beta] (0|\tau)} ~,
\label{sigmaf}
\ee
where the ($F$--dependent) matrix $\kappa$ is given by
\be
\kappa^{il} = {1\over 2} \eta^{ij} {(A^{-1})^l_{\,\ j}} ~.
\label{kap}
\ee
For fixed characteristic, $\sigma$--functions corresponding to different 
Weierstrass polynomials are related by
\be
\sigma^{{\hat F}}[\vec\alpha,\vec\beta] (\vec v) = \exp{\left(\med \sum_{i,j} 
d_{ij} v_i v_j\right)} ~\sigma^{F}[\vec\alpha,\vec\beta] (\vec v) ~.
\label{sigdiff}
\ee
Notice that all the ingredients defined so far enter in Eq.(\ref{sigmaf}).
As mentioned above, the $\sigma$--function is duality invariant; that is, 
invariant under the action of the modular group ${\rm Sp}(2g, \dZ)$. The 
exponential factor in (\ref{sigmaf}) is choosen in such a way that it 
cancels the duality transformation properties of the quotient of 
$\Theta$--functions.

The hyperelliptic {\em Kleinian functions} are nothing but derivatives of
the $\sigma$--function:
\begin{subequation}[alph]
\be 
\zeta_j^{F}[\vec\alpha,\vec\beta] (\vec v) = {\partial \ln \,
\sigma^{F}[\vec\alpha,\vec\beta] (\vec v) \over \partial v_j} ~,
\ee
\be
\wp_{ij}^{F}[\vec\alpha,\vec\beta] (\vec v) = - {\partial^2 \ln \, 
\sigma^{F} [\vec\alpha,\vec\beta] (\vec v) \over\partial  v_i \partial v_j} ~,
\ee
\end{subequation}
further indices denoting higher derivatives with respect to $v_k$. These
functions satisfy differential equations that generalize Weierstrass' cubic 
relation \cite{BuEnLe}. Furthermore, Kleinian functions that differ in their
generating Weierstrass polynomial are related by
\be
\wp^{{\hat F}}_{ij}(\vec v) = \wp^{F}_{ij}(\vec v) - d_{ij} ~,
\label{weierf}
\ee
regardless of the characteristic. It was already shown by Bolza \cite{Bo4}
that
$$
\sum_{i,j} \wp_{ij}^F[\vec\alpha,\vec\beta](0) x_1^{g-i} x_2^{g-j} =
{F(x_1,x_2) - Q(x_1)R(x_2) - Q(x_2) R(x_1)  \over 4 (x_1-x_2)^2} ~,
$$
where $F$ is an arbitrary Weierstrass polynomial, and $[\vec\alpha,\vec\beta]$ 
is the characteristic associated to the factorization $y^2=Q(x)R(x)$. Thus, 
it is evident that $\wp_{ij}^F[\vec\alpha,\vec\beta](0)$ vanishes for the 
Weierstrass polynomial (\ref{bolzapol}). This is precisely the quadratic
contribution to the $\sigma$--function. Consequently, as an outcome of this
approach, we can fully characterize the blowup function of $SU(N)$ twisted 
${\cal N}=2$ super Yang--Mills theories \cite{EdGRMar}:

~

\noindent $\triangleright$
The blowup function of $SU(N)$ Donaldson--Witten theory, in the absence of 
magnetic flux, is a hyperelliptic $\sigma$-function with Weierstrass 
polynomial $F(x_1,x_2) = 2 (P_N(x_1) P_N(x_2) - 4 \Lambda^{2N})$ and
characteristic $(\vec\Delta,\vec 0)$,
\be
\tau(t_k|u_j) = \sigma^{F}[\vec\Delta,\vec 0](it_{k+1}) ~.
\label{blowupf}
\ee
after identification of the Jacobian coordinates $v_k$ with the {\em times}
$it_{k+1}$ (the reasons behind the latter name will be clear shortly).

~

\noindent An immediate corollary of this result is the following:

~

\noindent $\triangleright$
The contact terms ${\cal T}_{k+1,l+1}$ are given by
\be
{\cal T}_{k+1,l+1} = \kappa^{k,l} = - {1\over 8 \pi i}{\partial u_{l+1}
\over \partial a^i}\oint_{A^i}P_k'(x)P_N(x){dx \over y} ~,
\label{cont}
\ee
where $\kappa$ is the matrix introduced in (\ref{kap}), and we have used
the explicit expression for $dr^k$ given in (\ref{drzero}).

~

In the remaining sections we will extract interesting consequences 
from these formulas. We will show, in particular, that this expression for 
the contact terms turns out to be very useful when considering the case of 
manifolds of simple type, where the only nonvanishing contributions to the 
$u$-plane integral come from those points of $\cM$ with maximal number of 
mutually local massless monopoles. The coincidence of (\ref{cont}) with the 
standard result (\ref{confin}), can be shown by means of the Whitham equations
(the Renormalization Group equations in the formalism of Ref.\cite{GoMaMiMo})
that express the derivatives of the moduli with respect to $T_n$ --closely 
related to the contact terms--, in terms of $A$-periods of a different basis
of Abelian differentials of the second kind \cite{EdGRMar}.

\section{The blowup function and the KdV hierarchy}

In order to show our main result, \ie that the blowup function satisfies 
the differential equations of the KdV hierarchy, we shall first analyze the
effect of special linear transformations on it. To this end, it is
extremely useful to introduce a tricky {\em symbolic} notation --widely
used by Baker in the nineteenth century \cite{Ba1,Ba2,Ba3}-- for the 
geometrical data of the Seiberg--Witten solution. The hyperelliptic curve 
$\Sigma$ can be written as
\be
y^2 = (\alpha_1 + \alpha_2 x)^{2g+2} ~,
\label{symb}
\ee
where, of course, $\alpha_1$ and $\alpha_2$ are not complex numbers. The only
meaningful object is the combination thereof
\be
\lambda_p = {2g+2 \choose p} \alpha_1^{2g+2-p} \alpha_2^p ~,
\label{symboltrans}
\ee
that renders (\ref{symb}) equal to (\ref{thecurve}). This notation allows us
to introduce a quite interesting Weierstrass polynomial, the so-called 
$(g+1)$--polar of the hyperelliptic curve,
\beqa
\tilde F (x_1,x_2) & = & 2 (\alpha_1 + \alpha_2 x_1)^{g+1} (\alpha_1 + 
\alpha_2 x_2)^{g+1} \nonumber \\
& = & 2\sum_{p,q=1}^{g+1} {{g+1 \choose p}{g+1 \choose q} \over {2g+2 
\choose p+q}} ~\lambda_{p+q} ~x_1^p x_2^q ~,
\label{polar}
\eeqa
which is {\em covariant} with respect to an ${\rm Sl}(2,\dR)$ transformation
of the $x$-coordinates. Indeed,
\be
\tilde F (x_1,x_2) = (c + d t_1)^{-g-1} (c + d t_2)^{-g-1} \tilde F
(t_1,t_2) ~,
\label{fdoscov}
\ee
under $x = (a + bt)/(c + dt)$, ~$bc-ad =1$. The hyperelliptic curve, in
turn, becomes
\be
Y^2 = (\beta_1 + \beta_2 t)^{2g+2} = \sum_{i=0}^{2g+2} \widehat \lambda_i
t^i ~,
\label{canonical}
\ee
where $Y= (c + dt)^{g+1}~y$, ~$\beta_1 = c \alpha_1 + a \alpha_2$, ~$\beta_2
= d \alpha_1 + b \alpha_2$. It is clear that, by these means, one can always
drive the hyperelliptic curve into its canonical form, ~$\widehat 
\lambda_{2g+2} = \beta_2^{2g+2}=0$, ~$\widehat \lambda_{2g+1} = \beta_1 
\beta_2^{2g+1} = 4$. The Abelian differentials of the first kind transform 
linearly,
\be
x^{g-i} {dx \over y} = (a + bt)^{g-i} (c+ dt)^{i-1} {dt \over Y} ~~~
\Rightarrow ~~~ dv_i(x) = \Lambda_i^{\,m} d\hat v_m (t) ~,
\label{abdiffirst}
\ee
where $\Lambda_i^{\, m}$ is an invertible matrix. Their periods get modified
accordingly, $A^i_{\,j} = \widehat A^i_{\,m}\Lambda_j^{\, m}$, ~$B_{ij} =
\widehat B_{im}\Lambda_j^{\,m}$ so that
\be
\tau_{ij} = \widehat \tau_{ij} ~.
\label{coupl}
\ee
Now, taking into account the covariance of $\tilde F$, the transformation 
properties of the $\eta$--periods can be extracted --even without knowing 
the corresponding basis of Abelian differentials of the second kind--,
\be
\widehat \eta ^{ij} = \Lambda_k^{\,\, i} \eta^{kj} ~.
\label{etatrans}
\ee
Thus, a linear transformation of the {\em times}
\be
\widehat v_l =(\Lambda^{-1})_l^{\, m} v_m ~,
\label{lintrevti}
\ee
yields
\be
\sigma^{\tilde F} [\vec\alpha,\vec\beta](v_l)_{(x,y)} = \sigma^{\tilde F} 
[\vec\alpha,\vec\beta](\widehat v_l)_{(t,Y)} ~,
\label{yields}
\ee
where $\tilde F$ denotes here the polar associated to the corresponding 
curves. After substituting $v_l =\Lambda_l^{\, m} \widehat v_m$, $\sigma^F 
[\vec\alpha,\vec\beta](v_l)_{(x,y)}$ satisfies the same differential 
equations than $\sigma^F [\vec\alpha,\vec\beta ](\widehat v_l)_{(t,Y)}$
with respect to the hatted times.

~

There is a third Weierstrass polynomial that plays an important r\^ole in
our proof. Let us call it $\hat F$. It was introduced by Baker \cite{Ba1} a
long time ago and revisited recently in Ref.\cite{BuEnLe}. It reads:
\be
\hat F (x_1,x_2) = 2 \lambda_{2g+2} x_1^{g+1} x_2^{g+2} + \sum_{i=0}^g
x_1^i x_2^i (2\lambda_{2i} + \lambda_{2i + 1} (x_1 + x_2)) ~,
\label{fvictor}
\ee
and the corresponding basis of Abelian differentials of the second kind is
\be
dr^j = \sum_{k=g+1-j}^{g+j} (k+j-g) ~\lambda_{k-j+g+2} ~{x^k dx \over 4y} ~.
\label{basisone}
\ee
As already mentioned above, hyperelliptic Kleinian functions satisfy 
differential equations which generalize those of the elliptic case like,
for example, Weierstrass' relation $(\wp'(u))^2 = 4 \wp(u)^3 - g_2 \wp(u) -
g_3$. They were first studied by Baker in the case of genus two \cite{Ba3},
and a generalization of his construction has been recently worked out
\cite{BuEnLe}. The relevant differential equations are rather implicit.
For the derivatives of $\wp^{\hat F}_{11}$ one can, however, write an explicit 
equation for arbitrary genus:
\beqa
& & \wp^{\hat F}_{111i} = (6 \wp^{\hat F}_{11} + \lambda_{2g}) \wp^{\hat 
F}_{1i} + {1\over 4} \lambda_{2g+1} (6 \wp^{\hat F}_{i+1,1} - 2 \wp^{\hat 
F}_{i2} + {1\over 2} \delta_{i1} \lambda_{2g-1}) \nonumber \\
& & ~~~+ {1\over 2} \lambda_{2g+2} (6 \wp^{\hat F}_{i+2,1} - 6 \wp^{\hat 
F}_{i+1,2} + 2 \wp^{\hat F}_{i3} - \delta_{i1} \lambda_{2g-2} - {1\over 2} 
\delta_{i2} \lambda_{2g-3}) ~. ~~~~~~~
\label{urkdv}
\eeqa
This equation, being of second order, is independent of the characteristic.
It only shows up in the choice of initial conditions. A change in the
Weierstrass polynomial amounts, after (\ref{weierf}), to a $v$-independent 
shift. Let us now drive the curve to its canonical form by means of an ${\rm
Sl}(2,\dR)$ transformation. The equation above becomes
\be
\wp^{\hat F}_{111i} = (6 \wp^{\hat F}_{11} + \widehat \lambda_{2g}) \wp^{\hat
F}_{1i} + 6 \wp^{\hat F}_{i+1,1} - 2 \wp^{\hat F}_{2i} + {1\over 2} 
\delta_{i1} \widehat \lambda_{2g-1} ~,
\label{wei}
\ee
where $\hat F$ is the polar associated to the canonical curve. It is now
easy to show that eq.(\ref{wei}) implies that the hyperelliptic Kleinian
functions satisfy the equations of the KdV hierarchy \cite{BuEnLe}. Indeed, 
take ${\cal U} = 2 \wp_{11} + {1\over 6} \widehat \lambda_{2g}$, put $x 
\equiv v_1$, and let $t_i = v_i$ be the higher evolution times, so that
\be
{\partial {\cal U} \over \partial t_2} = {1 \over 4} {\cal U}'''- {3 \over
2} {\cal U} {\cal U}' ~,
\label{eslaec}
\ee
where $'$ denotes derivatives with respect to $x$. This is precisely the 
KdV equation. In fact, ${\cal U}$ is a $g$-gap solution of the KdV hierarchy.
To see this, recall that the higher evolution equations of the hierarchy are
(for a review, see Appendix A of Ref.\cite{DFGiZJ}),
\be
{\partial {\cal U} \over \partial t_i} = R_i'({\cal U}, {\cal U}', \cdots) ~,
\,\,\,\,\,\,\,\,\ i\ge 3 ~,
\ee
where the functions in the right hand side are defined recursively,
\be
R_{i+1}'={1 \over 4} R_i'''- ({\cal U}+ {\widehat \lambda_{2g} \over
12})R_i'-{1\over 2}{\cal U}'R_i ~.
\ee
The blowup function (\ref{blowupf}) can be finally written as
\be
\tau(v_m=\Lambda_m^{\,\,\ l}\widehat v_l|{\cal O}_i) = {\rm 
e}^{\sum_{ij}c_{ij}\widehat v_i \widehat v_j} \sigma^{\hat F}[\vec\Delta,
\vec 0](\widehat v_l)_{(t,Y)} ~,
\ee
where $\Lambda$ is the appropriate transformation yielding the hyperelliptic
curve canonical, and the $c_{ij}$ are constants depending on the parameters
of the ${\rm Sl}(2,\dR)$ transformation and the moduli of the curve, that
can be computed explicitly by comparison of $\sigma$--functions defined for
different Weierstrass polynomials. We have thus arrived to the following 
result:
\be
{\cal U} = - 2 {\partial^2 \log\tau \over \partial \widehat v_1^2} + 4 c_{11}
+ {1 \over  6} \widehat \lambda_{2g} ~,
\ee
is a $g$-gap solution of the KdV hierarchy. In other words, the blowup 
function is --up to a redefinition of the evolution times and a constant 
shift--, a $\tau$--function of the KdV hierarchy.

The blowup function appears in the generating function of the correlators
involving the exceptional divisor. Thus, a corollary of the above result
is that these correlation functions on the manifold $\widehat X$ are governed
by the KdV hierarchy, and they have as initial conditions the generating
function of the original manifold $X$. It is intriguing that the differential
equations turn out to be essentially the same than those governing 
the correlation functions of two-dimensional topological gravity \cite{Wi91},
though the blowup function --being a $g$-gap solution-- lies far appart of 
these correlation functions in the space of solutions of the KdV hierarchy.
Finally, it is now clear that the differential equation originally used in 
the elliptic case \cite{FiSt}, can be understood, under the light of these
results, as the reduction of the KdV equation.

\section{The blowup function for manifolds of simple type}

We have already mentioned that the whole nontrivial contribution to the 
$u$-plane integral for manifolds of simple type comes from $N$ maximal
singularities of ${\cal M}$. These singularities, where $N-1$ mutually local
monopoles get massless, are also known as ${\cal N}=1$ points because they
are the confining vacua after breaking ${\cal N}=2$ down to ${\cal N}=1$.
The curve $\Sigma$ can be described in the vicinity of one of these points 
by Chebyshev polynomials (we set $\Lambda = 1$) \cite{DoSh}
\be
P_N(x) = 2 \cos \bigl( N \arccos {x \over 2} \bigr) ~,
\label{chevy}
\ee
and the other ${\cal N}=1$ points are obtained from the former by means of
the ${\dZ}_N$ symmetry of the theory. From now on we will focus on this 
${\cal N}=1$ point. There, the branch points of the curve become (single) 
$e_1=-e_{2g+2}=2$, and (double) $e_{2k}=e_{2k+1}=\widehat \phi_k = 2 \cos 
\widehat \theta_k$, ~$\widehat \theta_k = (\pi k/N)$. The values of the 
Casimirs are given by the elementary symmetric polynomials of the 
eigenvalues $2 \cos\theta_i$, ~$\theta_i = \pi (i-1/2)/N$. For example,
$u_2 = N$, ~$u_3 = 0$, ~$u_4={N\over 2}(3-N)$, ~etc. The $B$--cycles 
surround the points $\widehat \phi_i$ clockwise, while the $A$--cycles 
become curves going from $\widehat \phi_i$ to $2$ on the upper sheet and 
returning to $\widehat \phi_i$ on the lower sheet. The hyperelliptic curve 
becomes
\be
y ={\sqrt {x^2-4}} ~\prod_{k=1}^g (x-\widehat \phi_k) ~.
\label{eslay}
\ee
Consider now the normalized {\em magnetic} holomorphic differentials,
\be
\omega^j = (B^{-1})^{k j} dv_k = - {2i \sin\widehat \theta_j \over {\sqrt
{x^2-4}} ~(x-\widehat \phi_j)} ~,
\label{magholdiff}
\ee
that is, the canonical basis of Abelian differentials of the first kind
with respect to the $B$--cycles,
\be
{1\over 2\pi i} \oint_{B_i} \omega^j = - {\rm res}_{x =\widehat \phi_i} 
\omega^j = \delta^j_i ~.
\label{bcyc}
\ee
We can explicitely compute the derivatives of the moduli with respect to
the dual coordinates,
\be
{\partial u_{\ell+1} \over \partial a_{D,m}} = 2i(-1)^{\ell} \sin\widehat 
\theta_m ~E_{\ell-1} (\widehat \phi_{p\not=m}) ~,
\label{ders}
\ee
where $E_0=1$, and $E_j = \sum_{i_1 < \cdots < i_j} x_{i_1} \cdots x_{i_j}$
are the elementary symmetric polynomials of degree $j$. Near the ${\cal N}=1$
points, the diagonal components of the {\em magnetic} couplings diverge, but
the off-diagonal components are finite. The leading terms of the off-diagonal
components have been investigated in Ref.\cite{DoSh}, where an implicit 
expression for them was proposed in terms of an integral involving a
scaling trajectory. In the framework of the Whitham hierarchy approach to
four-dimensional ${\cal N}=2$ supersymmetric gauge theories, some nontrivial
constraints arise on those terms, and an explicit expression satisfying the 
constraints was proposed \cite{EdMa}. We will now derive a very simple 
expression for the leading terms of the off-diagonal couplings. From the 
above considerations, it follows that
\be
\tau_D^{k \ell} = {1\over \pi i} \int_{\widehat \phi_k}^2 \omega^\ell = {1 
\over \pi i} \log {\gamma_\ell -\gamma_k \over \gamma_\ell+\gamma_k} ~, 
\,\,\,\,\,\,\,\,\ k<\ell ~,
\label{magcoup}
\ee
where 
\be
\gamma_j = -i {\sqrt {\widehat \phi_j -2 \over \widehat \phi_j +2}} =
\tan {\pi j \over 2 N} ~.
\label{gammas}
\ee
This expression agrees with that conjectured in Ref.\cite{EdMa}, as it was
proved very recently by Braden and Marshakov \cite{BrMa}.

In order to compute the contribution of the ${\cal N}=1$ points to the
blowup function of simple type manifolds, we must first perform a duality
transformation to the {\em magnetic} coordinates. Being duality invariant,
the blowup function remains the same, except for the fact that everything
has to be replaced by its dual. In particular, the characteristic dual to 
$[\vec\Delta,\vec 0]$ is $[\vec 0,\vec\Delta]$. In order to compute the
contact terms, it is extremely useful to use the expression we derived
before, say (\ref{cont}), adapted to the dual frame. We just have to 
compute the $B$--periods of the Abelian differentials of the second kind 
(\ref{drzero}) at the ${\cal N}=1$ point. After the change of variables 
$x=2 \cos\theta$,
\be
dr^\ell = i P'_\ell(\theta) \cot N\theta ~\sin \theta ~d\theta ~,
\label{druno}
\ee
and their $B$--periods simply read
\be
\eta^\ell_{\,\, k} = {\rm res}_{\theta =\hat \theta_k} dr^{\ell} = {i \over 
N} P'_\ell(\widehat \phi_k) \sin \widehat \theta_k ~.
\label{etanuno}
\ee
The contact terms are then given by
\be
{\cal T}_{k,\ell} = {i \over 2N} P'_{k-1}(\widehat \phi_m) \sin \widehat
\theta_m ~{\partial u_{\ell} \over \partial  a_{D,m}} ~.
\label{conti}
\ee
We can now write the resulting expression for the blowup function. Notice
first that the dual $\Theta$--function vanishes at the ${\cal N}=1$ point. 
However, after quotienting by $\Theta[\vec 0,\vec\Delta](0|\tau_D)$, we get 
a finite result. At the end of the day, the blowup function corresponding
to simple type manifolds is given by
\beqa
\tau (t_i) & = & ~\left[\sum_{s_p=\pm 1} \prod_{p<q} \Bigl( {\gamma_q
-\gamma_p \over \gamma_q + \gamma_p}\Bigr)^{s_p s_q/2} \right]^{-1} \exp 
\Biggl\{ -\sum_{k,\ell} t_{k} t_\ell {\cal T}_{k,\ell} \Biggr\} \nonumber \\
& & ~~~\cdot \sum_{s_j=\pm 1} \prod_{p<q} \Bigl( {\gamma_q
-\gamma_p \over \gamma_q + \gamma_p}\Bigr)^{s_p s_q/2} \exp \Bigl\{
\sum_{l=2}^N{{i s_j t_l \over 2} {\partial u_{l} \over \partial
a_{D,j}}}\Bigr\} ~,
\label{finalblow}
\eeqa
with ${\cal T}_{k,\ell}$ given above. It is, after a linear transformation 
of the times, a $\tau$ function for an $(N-1)$--soliton solution of the 
underlying KdV hierarchy \cite{EdGRMar}. This is a simple consequence of 
the fact that quasi-periodic solutions of the KdV hierarchy become 
multisoliton solutions in the limit of maximal degeneracy of the underlying 
Riemann surface \cite{BBEIM} (see also Ref.\cite{BrMa} for recent progress 
in this direction). Let us finally point out that, regardless of its apparently
involved expression, the blowup function for simple type manifolds turns out
to be, on general grounds, simplified. For example, in the case of $SU(3)$,
eq.(\ref{finalblow}) reads
\be
\tau_{SU(3)} (t_2,t_3) = {1\over 3}{\rm e}^{-{1\over 2}t_2^2 -t_3^2}
\Bigl\{ \cosh ( {\sqrt 3}t_2) + 2 \cosh ( {\sqrt 3}t_3) \Bigr\} ~.
\label{sutst}
\ee
This fact was already observed in the  elliptic case \cite{FiSt}, and has
to do with the degeneration of hyperelliptic functions into trigonometric ones.

An important consistency check of our expression for the blowup function can
be made by considering the explicit expression of the Donaldson--Witten 
generating function for manifolds of simple type \cite{MaMo2}, which can be 
trivially extended to include more general descent operators \cite{EdGRMar}:
\beqa
& & Z(p_k,f_k,S)_X^{{\cal N}=1} = ~\alpha^{\chi} \beta^{\sigma} \sum_{x_j} 
\bigl( \prod_{j=1}^{N-1} SW(x_j) \bigr) \prod_{j<k} \Bigl( {\gamma_k - 
\gamma_j \over \gamma_j + \gamma_k} \Bigr)^{-{(x_j,x_k)/ 2}}~~ \nonumber \\ 
& & ~~~~~\cdot ~\exp\Bigl\{ \sum_{k=2}^N \Bigl( p_k u_k - 
{i\over 2} f_k {\partial u_{k} \over \partial a_{D,j}}(S,x_j)\Bigr) + S^2 
\sum_{k,l}f_kf_l {\cT}_{k,l} \Bigr\} ~,
\label{sundw}
\eeqa
where the values of the $B$--periods and contact terms are those given in 
(\ref{ders}) and (\ref{conti}), and $\alpha$ and $\beta$ are universal 
constants that only depend on $N$. Only the contribution of one of the
${\cal N}=1$ points is recorded in the equation above, those of the other 
points following from ${\dZ}_N$ symmetry. For each $i=1, \cdots, N-1$, the 
sum over $x_i$ is over all the Seiberg--Witten basic classes of the manifold
$X$ \cite{Wi94}, whose Seiberg--Witten invariants are denoted by $SW(x_j)$.
$(\,\  , \,\ )$  denotes the product in (co)homology. After a blowup, every 
basic class $x$ of $X$ leads to basic classes $x \pm B$ in $\widehat X$, 
where latter $x$ really denotes the pullback to $\widehat X$ of the basic 
class of $X$. The Seiberg--Witten invarians are $SW(x\pm B) = SW(x)$. 
If we now consider $Z(p_k, f_k,S)_{\widehat X}^{{\cal N}=1}$, 
we will have to substitute $x_i \rightarrow x_i +s_iB$ in (\ref{sundw}), 
with $s_i =\pm 1$. The sum over basic classes of $\widehat X$ factorizes 
into a sum over the $x_i$ and a sum over the $s_i$. Taking into account 
that $(x,B)=0$ for any cohomology class $x$ pulled back from $X$ to $\widehat
X$, and that $B^2=-1$, eq.(\ref{sundw}) gets an extra factor under blowup
that agrees with (\ref{finalblow}) \cite{EdGRMar}. This is a crucial
nontrivial consistency check of the whole set of results presented in this 
talk, since, when using (\ref{sundw}), we have to rely on properties of the 
Seiberg--Witten invariants, while the blowup function (\ref{finalblow}) was 
derived by means of the $u$-plane integral.

\section{Final remarks}

An important aspect of blowup functions \cite{MoWi,MaMo,MaMo2} is that they 
must admit an expansion of the form (\ref{infser}), whose coefficients are 
polynomials in the local observables of the topological field theory.
In the case of $SU(2)$, this is rather explicit from the theory of elliptic
functions. As recently shown in Ref.\cite{EdGRMar}, it turns out that an
elegant and powerful recursive method to perform this expansion in the case 
of $SU(N)$ --up to arbitrary order in the {\em times}--, follows from the 
theory of hyperelliptic Kleinian functions \cite{Bo1,Bo2,Bo3,Bo4}. For 
example, in the case of $SU(3)$, the outcoming result is \cite{EdGRMar}
\beqa
& & \tau_{SU(3)}(t_i|u_i) = ~1 - {\Lambda^6\over 12} \Bigl[ u_2 t_3^4 + 6
t_2^2 t_3^2 \Bigr] - {\Lambda^6\over 360} \Big[ 3 t_2^6 - 15 u_2 t_2^4 t_3^2 -
60 u_3 t_2^3 t_3^3  \nonumber \\
& & ~~~~~ - 15 u_2^2 t_2^2 t_3^4 - 12 u_2 u_3 t_2 t_3^5 - u_2^3 t_3^6 + 3 
u_3^2 t_3^6 - 12 \Lambda^{6} t_3^6 \Bigr] + \cdots ~.
\eeqa
A further consistency check comes from duality invariance of the blowup 
function: this expansion coincides with (\ref{sutst}) when $u_2 = 3$ and 
$u_3 = 0$.

The theory of hyperelliptic Kleinian functions has shown to be an appropriate
framework to address many aspects of the blowup formulas and the $u$-plane 
integral, like contact terms and the relation with integrable hierarchies.
It would be very interesting to work out the details for theories including
massive hypermultiplets and/or other gauge groups. In massive theories, for
example, whose Whitham formulation was worked out in \cite{EdGRMaMa},
the magnetic flux through the class of the exceptional divisor turns out to be 
fixed by topological constraints, this giving a nonzero value of 
$\vec\beta$ in the blowup function \cite{MoWi,EdGRMar2}. 

Another direction to explore is the relation between the hyperelliptic
Kleinian functions and the theory of the prepotential. The blowup function
gives a natural set of Abelian differentials of the second kind
--which differs from the one given in \cite{GoMaMiMo}--, and we know
from general principles that such a set is one of the basic ingredients in
the construction of a Whitham hierarchy \cite{Kr}. It would be very
interesting to develop this relation in general, at least for hierarchies
associated to hyperelliptic curves. This would further clarify the
relations between blowup functions in generalizations of Donaldson--Witten
theory, and the construction  of Whitham hierarchies for supersymmetric  
${\cal N}=2$ theories.

\acknowledgements

We would like to thank Marcos Mari\~no and Javier Mas for gratifying
collaborations leading to the results discussed here. J.D.E. wishes 
to thank Henrik Aratyn and Alexander Sorin for their kind invitation to 
present these results in such an enthusiastic environment, and for their 
delightful hospitality. The work of J.D.E. has been supported by the 
Argentinian National Research Council (CONICET) and by a Fundaci\'on 
Antorchas grant under project number 13671/1-55.



\end{document}